\documentclass[aps,pra,twocolumn,showpacs,superscriptaddress]{revtex4-1}  
\usepackage{times,amsmath,amssymb,amsfonts,graphicx,color}
\usepackage{graphicx}  
\usepackage{dcolumn}   
\usepackage{bm}        
\usepackage{amssymb} 
\usepackage{color}

\begin{document}
\title{Minimal model for spontaneous { quantum } synchronization}
\author{Claudia Benedetti}
\email{claudia.benedetti@unimi.it}
\homepage{users.unimi.it/aqm}
\affiliation{Quantum Technology Lab, Dipartimento di 
Fisica, Universit\'a degli Studi di Milano, I-20133 Milano, Italy}
\author{Fernando Galve}
\email{fernando@ifisc.uib-csic.es}
\affiliation{IFISC (UIB-CSIC), Instituto de F\'{\i}sica 
Interdisciplinar y Sistemas Complejos,  Palma de Mallorca, Spain}
\author{Antonio Mandarino}
\email{mandarino@cft.edu.pl}
\affiliation{Quantum Technology Lab, Dipartimento di 
Fisica, Universit\'a degli Studi di Milano, I-20133 Milano, Italy}
\affiliation{Center for Theoretical Physics PAS, 02-668 Warsaw,
Poland}
\author{Matteo G. A. Paris}
\email{matteo.paris@fisica.unimi.it}
\homepage{users.unimi.it/aqm}
\affiliation{Quantum Technology Lab, Dipartimento di 
Fisica, Universit\'a degli Studi di Milano, I-20133 Milano, Italy}
\author{Roberta Zambrini}
\email{roberta@ifisc.uib-csic.es}
\affiliation{IFISC (UIB-CSIC), Instituto de
F\'{\i}sica Interdisciplinar y 
Sistemas Complejos, Palma de Mallorca, Spain}
\date{\today}
\begin{abstract}
We show the emergence of spontaneous synchronization between a pair of
detuned quantum oscillators within a harmonic network.  Our model does
not involve any nonlinearity, driving or external dissipation, thus
providing the simplest scenario for the occurrence of local coherent
dynamics in an extended harmonic system. A sufficient condition for
synchronization is established building upon the Rayleigh's normal modes
approach to vibrational systems. Our results show that mechanisms
favoring synchronization, { even between 
oscillators that are not directly coupled to each other}, are
transient energy depletion and cross-talk.  We also address the possible
build-up of quantum correlations during synchronization and show that
indeed entanglement may be generated in the detuned systems starting
from uncorrelated states and without any direct coupling between the 
two oscillators.
\end{abstract}
\maketitle
\section{Introduction}
Synchronization among dynamical systems is a widespread phenomenon which 
has been widely studied in the classical domain and, more recently, in 
the quantum regime \cite{SynchQ,sync2012,syncSR,fazio1}. Indeed, 
synchronization is a relevant phenomenon in several contexts, including 
physical, biological, chemical and social systems, and it 
has been also generalized to 
a large variety of  dynamical regimes,
from regular oscillations to chaotic evolutions 
\cite{stro93,osi07,Pikovsky,Strogatz,manb}.
These investigations have extended the definition of synchronization,
showing that mutual or directional coupling between inhomogeneous components,
are relevant for its occurrence, as well as
non-linearity, dissipation, noise, forcing, or time delay. Overall, 
synchronization emerges as a paradigmatic phenomenon in complex systems
\cite{Pikovsky}. 

{
The extension of the synchronization concept
to the quantum case is not straightforward, since dynamical trajectories of the observables are not well defined
and the very 
quantum properties of the canonical variables 
prevent the exact fulfillment of the classical
conditions defining synchronization \cite{fazio1,synch_chapter}. In turn, the appearance of
quantum synchronization has been proved to be different
from the appearance of coherence (as entanglement)
and a question arises on how, whether and when the two phenomena
may coexist. 
\par
In this framework,} the question about the necessary
ingredients to observe spontaneous synchronization in simpler dynamical
models was less explored. {
While in a classical setting, synchronization is mostly studied for 
nonlinear systems, in a recent work on the synchronization of
quantum fluctuations \cite{sync2012}, it was shown that it can actually
arise even in linear models,} solely due to dissipation, and it may
persist asymptotically for larger systems \cite{syncLalo,syncSR}.
\par
In this work we introduce and discuss a minimal model for the
emergence of synchronization, in systems with no external 
forcing, nonlinear effects, and dissipation. More specifically, 
we consider an isolated linear network of coupled harmonic 
oscillators and address the emergence
of coherent dynamics, i.e. synchronization, in the subsystem made 
of a pair of nodes.  In fact, harmonic network, besides being 
fundamental ingredients for modeling open quantum systems 
\cite{weiss-breuer}, are of interest for a broad
spectrum of topics ranging from consensus problems \cite{arenas} to
trapped ions \cite{ions}.   
\par
In general, harmonic networks models of extended environments, even in
the weak coupling limit, lead to a rather complex dissipation mechanisms
for the embedded subsystems \cite{ruggero,johannes,huelga,burghardt},
including several different spatial effects when the system is
multipartite \cite{CBSB,vega}. Our physical model involves 
a large isolated network of oscillators 
and within this description, we  analyze
dissipation-induced synchronization. Besides, 
we provide a sufficient condition for the emergence of synchronization 
between detuned nodes in the framework of the physics of vibrations 
and in the limit of the Rayleigh approximation. 
\par
In a second part of our work we analyze the strong coupling regime,
where the two nodes under investigation are strongly coupled to the 
rest of the network. This is still a linear model amenable to analytic 
solution. { Clearly, if one excites only one normal mode, 
some network nodes will oscillate synchronously, but this in not the case under study here.
We consider instead a generic initial 
condition exciting the two probes. Here,} the mechanisms governing the transport of energy 
as well as the possible scenarios for the emergence of synchronization 
are far to be trivial,
being typically limited to temporal transients and susceptible to
variations in couplings, inhomogeneities and boundary effects.  In 
particular, we present and discuss two different routes to  
synchronization mediated by the environment in a simple chain configuration. 
{ In addition, we analyze in some details the possible
build-up of quantum correlations during synchronization, showing that
entanglement may be generated in our detuned systems starting
from uncorrelated states and without any direct coupling between the 
two oscillators.}
\par
Indeed, synchronization in itself is defined using classical temporal averages
also for quantum systems. However, all the synchronization scenarios
mentioned above are not specific of classical systems, 
being not limited to first order moments.
In fact, quantum noise synchronization in presence of squeezing has been
already reported \cite{sync2012,syncSR}, upon 
considering local variances. We thus devote
the final Section to address the possible emergence of quantum
signatures, and analyze the dynamics of quantum correlations and the
possible build-up of mutual information and entanglement when starting
from uncorrelated product states.  The
possibility to generate quantum correlations through bosonic baths has
been already considered in the literature. In particular, ion chains
acting as reservoirs have been recently shown to mediate entanglement
between identical ions defects, when placed in one edge \cite{morigiY}
and also at a distance \cite{morigiEPL} while entanglement generation
via a heat bath could not be established between remote objects in Ref.
\cite{klesse}.  Here we take a further step extending previous analysis
to non-uniform systems (being the system components detuned), allowing
for strong system-environment coupling \cite{correa}, and establishing
the connections with a coherent (synchronous) dynamics
\cite{sync2012,fazio1,fazio2}.
\par
The paper is structured as follows. In Section \ref{sectMod} we 
introduce our model and establish notation. We also illustrate the
quantitative measure of synchronization used throughout the paper
and the Langevin equation governing the dynamics of the pair
of oscillators. In Section \ref{sectRay} we discuss a general
condition for synchronization and illustrate our results about
synchronization via weak dissipation or across the chain. In Section
\ref{sectRoutes} we show the results obtained in 
the strong coupling regime and discuss synchronization by coupling
to a common chain mode or by cross-talk. Finally, in Section \ref{secQ}
we show our results about the links between synchronization and the
build-up of quantum correlations for the oscillators coupled to a common
chain node or to the chain edges. Section \ref{out} closes the paper
with some concluding remarks.
\section{Dynamical model}\label{sectMod}
We consider a large network of $M$ coupled harmonic oscillators of unit mass
described by the Hamiltonian ($\hbar=1$)
$$H_{E}=\sum_{j=1}^M \frac{1}{2}(P_j^2+\Omega_0^2 X_j^2)+\sum_{j,k=1}^M
A_{jk}( X_j-X_k)^2\,.$$ Besides, we address a system of two detuned oscillators, 
coupled to a pair of nodes of the network with strength $K$,
and coupled between them with strength $\lambda$. 
The dynamics of the overall system is governed by the Hamiltonian
$H=H_S+H_{E}+H_{I}$, with 
\begin{eqnarray}
H_S&=&\frac{1}{2}(p_1^2+p_2^2)+\frac{1}{2} (\omega_1^2 x_1^2+\omega_2^2
x_2^2)+\frac{\lambda}{2} (x_1-x_2)^2\\ H_I &=&K(x_1\,X_m+x_2\,X_n),
\end{eqnarray}
where $n,m \in [1,M]$ are the positions within the network, where the two
detuned oscillators are plugged.
The  canonical operators for the oscillators in the chain are denoted 
by capital letters $[X_j,P_k]=i\delta_{jk}$, $[X_j,X_k]=0$, 
$[P_j,P_k]=0$, $j,k=1,M$ whereas 
$[x_j,p_k] = i \delta_{jk}$, $[x_j,x_k] = 0$, $[p_j,p_k] = 0$, 
$j,k=1,2$ 
are the operators for the two detuned oscillators (at frequencies
$\omega_j$, $j=1,2$). The
(common) natural frequency of the oscillators in the network is denoted by
$\Omega_0$ and the matrix $A_{jk}$ contains information about their
couplings. 
\par
In the limit of  $M\rightarrow \infty$ and decoupled oscillators, i.e.
$\lambda=0$, this is a well-known framework for open quantum systems
\cite{weiss-breuer} and can be used to microscopically derive
generalized Langevin equations for the reduced system dynamics
\cite{hanggi}. In this framework, one considers a set of independent degrees of
freedom in the environment---the environmental normal modes $Q_n$---and
assumes a certain spectral density, encoding the form of the coupling
between system and environment as well as the spectral distribution of
the latter. However, one can go beyond phenomenological assumptions and
derive the spectral density associated to more complex configurations of
coupled HOs, constituting different kind of finite networks
\cite{rubin,burghardt,ruggero,johannes}. 
The case of a homogeneous chain, i.e. $A_{jk} = g \delta_{|j-k|,1}$, is
particularly interesting because it allows (i) to reproduce an Ohmic
dissipation \cite{rubin} and (ii) to have a clear picture of the
transport dynamics.  On the other hand, increasing the environment
complexity allows to engineer arbitrarily complex spectral densities, as
in Refs.\cite{burghardt,ruggero,johannes}, exhibiting non-Markovian
effects \cite{huelga,NMrev}. 
\subsection{Synchronization}
Mutual synchronization arises when, in spite of their detuning,  
the pair of oscillators starts to oscillate
coherently, at a common frequency. A quantitative estimation 
of synchronization comes from a Pearson's correlation
among two time dependent functions $f,g$, namely
\begin{equation}
 \mathcal{C}_{f,g}(t,\Delta t)=\overline{ \delta f \delta
 g}/\sqrt{\overline{  \delta f^2} ~\overline{\delta g^2} }
\end{equation}
where the bar stands for a time average
$$\overline{f}=\int_{t}^{t+\Delta t}dt'f(t')$$ within a 
time window $\Delta t$ and $\delta f=f-\overline{f}$.  
This is an indicator measuring the
presence of dynamical synchronization between either classical
trajectories \cite{Pikovsky} or quantum systems\cite{sync2012}
characterized by average positions, variances, and, possibly 
higher order moments.  Other indicators of synchronization
consider different forms of correlations between the nodes as in Refs.
\cite{fazio1,fazio2}.
\par
As recently reported \cite{sync2012,syncSR}, a system of two (or more)
HOs {\em weakly} dissipating into an {infinitely} large thermal bath
($M\rightarrow \infty$) displays synchronous dynamics  when one normal
mode is more protected against dissipation than the other(s)
\cite{CBSB}.  In other words, synchronization emerges when all but one
modes are largely damped and the dynamics is then governed by the
eigenfrequency of the most robust mode \cite{syncLalo,syncSR}.  The
general condition derived for synchronization in the presence of a
weakly coupled and infinite bath is indeed the presence of a gap between
the damping rate of the two least damped modes of the system
\cite{syncSR}. For a finite environment and beyond weak coupling, the
scenario is more complex but richer and our first step is to identify a
similar mechanism for the emergence of synchronization, see Section 
\ref{sectRay}.
\subsection{Dynamics}
The dynamics of the subsystem of detuned oscillators, from now on
referred to as the {\em system}, is governed by a pair of 
integro-differential equations.
First and second order moments of the operators 
$(x_j$, $p_j)$, $j=1,2$ are sufficient to fully 
characterize Gaussian states and their dynamics.
We thus start by considering the average positions of the system
oscillators $\langle x_{1,2}(t)\rangle$,  whose dynamics is governed by
generalized quantum Langevin equations \cite{weiss-breuer,hanggi}. These
integro-differential equations depend on the structure and state of the
overall network, and for the normal modes $q_{1,2}$ of the system 
they read as follows 
(see Appendix \ref{app1})
\begin{eqnarray}\label{eq_av_NL}
 \langle\ddot{q}_{1}(t)\rangle  +\int_0^t dt'[\gamma_1(t-t')
 \langle\dot{q}_1(t')\rangle +\eta(t-t')\langle\dot{q}_2(t')\rangle] \nonumber\\
 +\Lambda^2_{1}\langle q_1(t)\rangle
 -\gamma_1(0)\langle q_1(t)\rangle -\eta(0)\langle q_2(t)\rangle=0 \nonumber 
\end{eqnarray}
Analogously, an equivalent equation is found for $\langle {q}_{2}\rangle
$ by replacing $1\leftrightarrow 2$.  Here the network features are
encoded in the time-dependent coefficients $\gamma_s$ ($s=1,2$) and
$\eta$, while $\Lambda_{1,2}$ are $H_S$  eigenfrequencies and we assume
$\langle q_{1,2}(0)\rangle=0$ as initial conditions.
\par
We emphasize that the system normal modes  $q_{1,2}$ diagonalize $H_S$
but they remain dynamically coupled through damping, due to the
interaction with the environment.  In fact, the damping kernel contains
different components; the first one is 
given by \begin{align}
\gamma_{s}(t-t')&=&\sum_{j=1}^M \frac{c_s^2(j)}
{\Omega_j^2}\cos[\Omega_j(t-t')]\Theta(t-t'),
\end{align}
with $s=1,2$, and governs the local damping at each node 
as well as the possible feedback from the boundaries 
of the finite chain. The second terms reads as follows
\begin{align}
\eta(t-t')&=&\sum_{j=1}^M \frac{c_{1}(j)c_{2}(j)}{\Omega_j^2}
\cos[\Omega_j(t-t')]\Theta(t-t')
\end{align}
and introduces cross effects in the friction, through the transmission 
of signals among the system components along the chain. For this reason the 
$\eta$ coefficient is symmetric. The mathematical expressions for the 
$c_s(j)$ and $\Omega_j$ coefficients are given in Appendix \ref{app1}.
The initial state for the network is the fundamental one ($T=0$),
being the initial energy excitation localized in the system 
oscillators only.
\section{A sufficient condition for synchronization}\label{sectRay}
The time non-local dissipation term in Eq.~(\ref{eq_av_NL}) may 
be approximated by a time local one, i.e. constant, damping, only 
in specific situations \cite{hanggi,Kimble}. This is usually 
the case when the system is weakly coupled to the rest of the 
network though, strictly speaking, each configuration of the network 
should be studied in details to understand whether and when 
a time local description is appropriate, at least during a 
transient time. If these conditions are fulfilled the dynamics 
of the system is described by a set of coupled differential equations
of the form
\begin{equation}\label{eq_Ray}
\langle {\bf \ddot x}\rangle+
G\langle {\bf \dot x}\rangle+A\langle {\bf x}\rangle =0\,,
\end{equation}
where $\bf x = {x_1,x_2}$ and $A$ and $G$ are time-independent matrices.
An interesting question, addressed earlier by Lord Rayleigh \cite{Ray} 
in the context of the vibration of structures  
\cite{Ray,vibrations} is whether normal modes may be individuated in spite 
of the presence of dissipation. The undamped dynamics follows from a 
superposition of normal modes obtained diagonalizing the 
stiffness matrix $G$ and the coupling one $A$ in 
Eq. (\ref{eq_Ray}), but the specific form of damping mines 
this description because, in general, $A$ and $G$ 
cannot be simultaneously diagonalized. 
\par
As a matter of fact \cite{Ray, Caugh}, {\em classical} 
normal modes \cite{clmode} are present if the matrices 
$A$ and $G$ commute. This leads to a simple description 
for the {\em independently} damped normal modes 
of the free dynamics 
(${\bf q}=(q_1,q_2)$).
Small deviations from the condition $[G,A]=0$ justify 
the Rayleigh's approximation of neglecting the non-diagonal 
components of $G$ in the basis of ${\bf q}$. This corresponds to the 
so-called {\em reduction method} \cite{Ray} of disregarding
out-of-diagonal terms of $G'=M^{-1}GM$, 
where $M=(\{q_1\},\{q_2\})$ diagonalizes 
$A$ ($A'=M^{-1}AM$), which is useful when dissipation is small.
The approximation is equivalent to neglect the small cross damping among
natural vibrations and, of course, the validity of this approach depends
on the relative size when comparing with self-dampings. An example is
shown in Ref.\cite{Galve} and based on a secular approximation.
The  model described in Eq.\eqref{eq_Ray}, simplified under  Rayleigh's
approximation, allows for a necessary and sufficient condition for
synchronization: a  pair of detuned oscillators embedded into a network
will synchronize if there exists a gap between the normal mode damping
rates $G'_{11}$ and $G'_{22}$. This condition is general for dissipation
in infinite baths \cite{syncSR} while for finite systems it is limited
to the transient where the average dynamics of the pair of oscillators
can be approximated by  Eq.~(\ref{eq_Ray}). Significant build-up of
synchronization requires the least damped mode to be suppressed and
this phenomenon should occur in a 
time scale of the order of the inverse of the larger damping  
\begin{equation}
 \tau_S^{-1}\sim\max(|G'_{11}|,|G'_{22}|).
\end{equation}
In the following Section, we consider a finite chain configuration
and provide and example of application of the above condition.
\subsection{Synchronization via weak dissipation} \label{secDiss}
Let us now consider a network made of a chain of $M$ oscillators 
homogeneous in frequency and couplings. The 
Hamiltonian is given by
$$H_E =\sum_{j=0}^M\frac{p_j^2}{2}+
\frac{1}{2}\Omega_0^2X_j^2+\frac{g}{2}(X_j-X_{j+1})^2\,.$$
The network acts as an environment for a system made of two
oscillators attached at one edge of the chain. The interaction
Hamiltonian reads as follows
\begin{equation}
\label{eqY}
H_I=K (x_1 + x_2) X_{1}. 
\end{equation}
In this configuration only $x_+=x_1+x_2$ is directly coupled to the 
chain and $x_\pm$ diagonalize the damping term, while the system 
Hamiltonian $H_S$ is diagonal in $q_{1,2}$.
\begin{figure}[h!]
 \centering
\includegraphics[width=0.9\columnwidth]{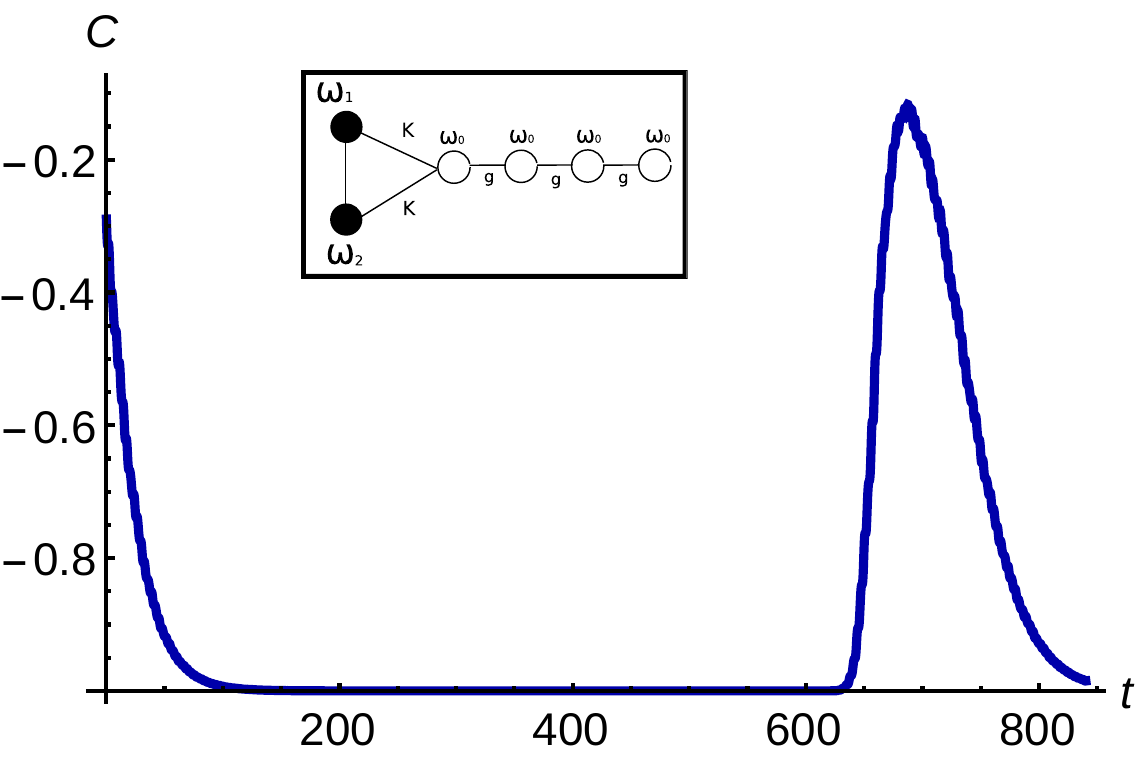}
\includegraphics[width=0.9\columnwidth]{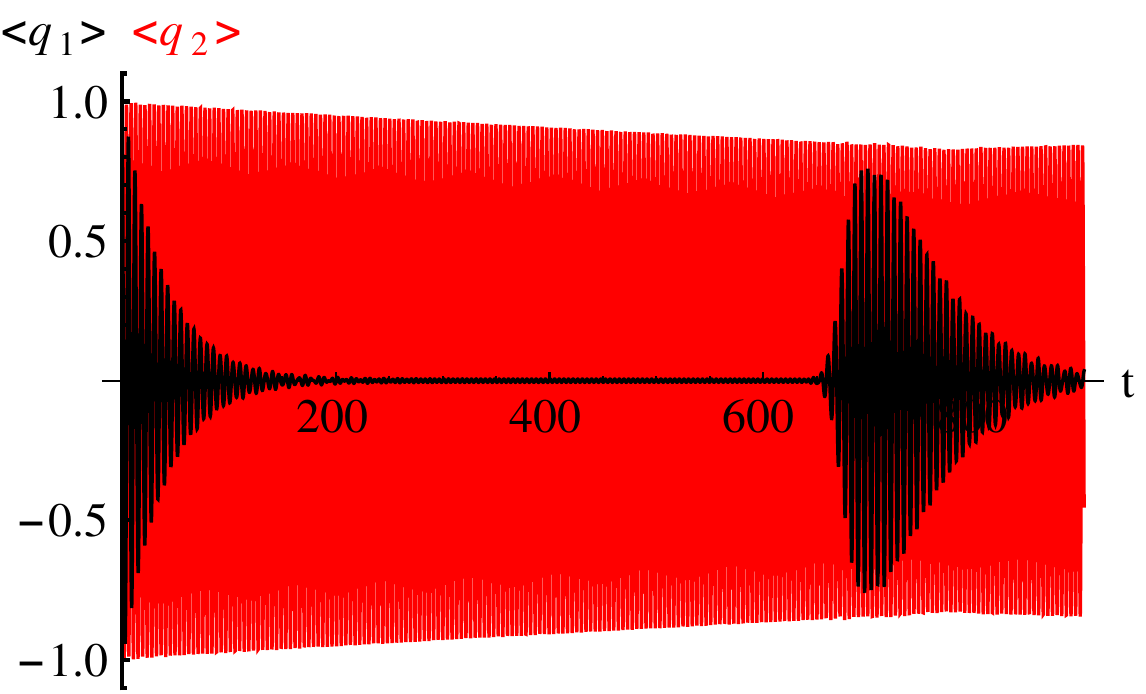}
\caption{Synchronization measure $C(t)$ and dynamics 
for $\omega_2/\omega_1=1.1$,  
$\lambda=0.5\, \omega_1^2$, $x_1(0)=0.14\,\omega_1^{-1/2}$; $x_2(0)=1.4\,{\omega_1}^{-1/2}$,
$M=300$, $\Omega_0=0.4\,\omega_1$, $g=1.2\,\omega_1^2$,
$K=0.2\,\omega_1^2$. In the lower panel the black line
denotes the function $\langle q_1(t) \rangle$ 
and the red one is for $\langle q_2(t) \rangle$. $\langle q_{1,2}\rangle$ are in units of $\omega_1^{-1/2}$ and time is in unit of $\omega_1^{-1}$. }
\label{f2}
\end{figure}
\par
In the limit of an infinitely large chain and vanishing 
local potential ($\Omega_0 \simeq 0$) the environment acts 
as an Ohmic bath and the ratio between the  damping rates is
\begin{equation}
 G'_{11}/G'_{22}\simeq \frac{1+\sin2\theta}{1-\sin2\theta}
\end{equation}
where the parameter $\theta$ depends upon the detuning and the coupling
$\lambda$ (see Appendix  \ref{app1}).  The sufficient condition for
transient synchronization is the presence of a gap between the normal
modes damping, as it happens for small detuning, i.e.
$|\omega_1-\omega_2|^2/ \lambda < 1$ in the case of 
Fig.~\ref{f2} where $\sin2\theta\simeq 1$.  
The synchronization measure $\mathcal{C}$ in this case shows
that perfect anti-synchronization is present up to the revival 
time $\tau_{R}\sim 2M/\omega_1$. 
If the coupling of one of the oscillators to the chain switches 
from attractive to repulsive, $K\,x_2X_1\rightarrow -K\,x_2X_1$ 
in Eq.(\ref{eqY}), then the quantity $x_-=x_1-x_2$ couples 
to the chain and synchronization instead of anti-synchronization arises.
\par
As a matter of fact, during the initial transient time 
finite size effects can be neglected and the energy of 
the two system's oscillators flows into the environmental 
chain \cite{en_flow}, leading to an effective dissipation 
into a common bath \cite{sync2012,syncSR}.  Therefore, the 
neat build-up of anti-synchronization of Fig.~\ref{f2} is 
consistent with the predicted phenomenon of Ref. 
\cite{sync2012}, where an infinite bath was considered.
Boundary effects cause a departure from the Ohmic dissipation (constant
damping in Eq.\ref{eq_av_NL}) leading to revivals.  Fig.\ref{f2} shows
that reflection from the boundary at $t=\tau_{R}$ actually deteriorates
the coherent dynamics between the pair of HOs, and similar results are
found when there are defects in the chain causing feedback effects at
shorter times.  This is accompanied by a regrowth in the oscillation
amplitude of the damped mode $q_1$. The loss of anti-synchronization is
indeed due to a common forcing toward synchronization due to the
feedback signal reflected at the edge of the chain. At later 
times ($t> \tau_{R}$), after a competition transient, 
anti-synchronization is restored under the effect of dissipation, 
as shown in Fig.\ref{f2}, lasting until feedback effects arise 
again at $t\sim 2\tau_{R}$. 
\subsection{Synchronization across the chain}
A natural question is what happens when moving 
the second oscillator through the chain
with system-environment interaction
\begin{equation}
H_{I}=K x_1 X_{1}+K x_2 X_{m} \label{eq_1m}~~~~~~~m\in [1,M].
\end{equation}
The dependence of dissipation with the distance ($m$) in the weak
coupling regime has been described elsewhere \cite{CBSB} for infinite
environment and a periodic transition between dissipation in common and
separate baths has been predicted.  The case under study differs due to
finite size effects: reflections from the boundaries and cross-talk
between the oscillators and feedback signals lead to a dynamics  which
strongly depends on the plugging distance $m$, and perfect
synchronization may arise or not just by moving the system components
from one site to the neighbor one.  Still, this sensitivity to the
plugging position is absent during a transient when $0\ll m\ll N$, i.e.
second oscillator far from the first one and form the edge of the chain.
More details are given in Appendix \ref{app2}.
\section{Synchronization in strong coupling regime}
\label{sectRoutes}
The mechanism of synchronization by dissipation is enabled by the
presence of coupling between the system oscillators (i.e. $\lambda\neq
0$) and it is consistent with results obtained for infinite environments
\cite{sync2012,syncSR}.  An interesting question  is the possibility to
synchronize detuned oscillators in the absence of a direct coupling between
them, i.e. $\lambda =0$, solely due to the mediating effect of the rest of
the network. This was actually shown to be possible for spins in
Ref.\cite{GLPlastina} but it does not occur for weakly coupled 
harmonic oscillators.
Indeed, for the oscillators pair attached to a common node, the
dissipation mechanism described above does not produce synchronization
in the weak coupling regime. Inspection of the master equation in
\cite{sync2012} shows that, even for long chains (large $\tau_{R}$) the
effective coupling induced by the bath (Lamb shift) is actually too
small to lead to significant synchronization before the system
thermalizes. On the other hand, a full system-bath model allows one to
address less explored strong dissipation regimes enabling new dynamical
scenarios for synchronization that are not present for weak coupling.
\subsection{Coupling to a common chain node}
\label{secSRedge}
We now consider a configuration where the system is 
attached to one edge of the environment chain, as in 
Eq.\eqref{eqY}, but now the two oscillators are 
uncoupled, $\lambda =0$. We allow for a frequency
detuning $\omega_1\neq\omega_2$ implying that $x_\pm$ 
are not the eigenmodes of $H_S$.  Up to the revival time 
$\tau_R$, the system oscillators will dissipate into 
the common environment and no synchronization is possible 
for weak coupling, i.e. $K\ll\omega_i^2$).
Under-damped detuned oscillations
characterize the dynamics at all times and the system components $x_{1}
,x_{2}$ remain incoherent. 
\par
\begin{figure}
\centering
\includegraphics[width=0.4\textwidth]{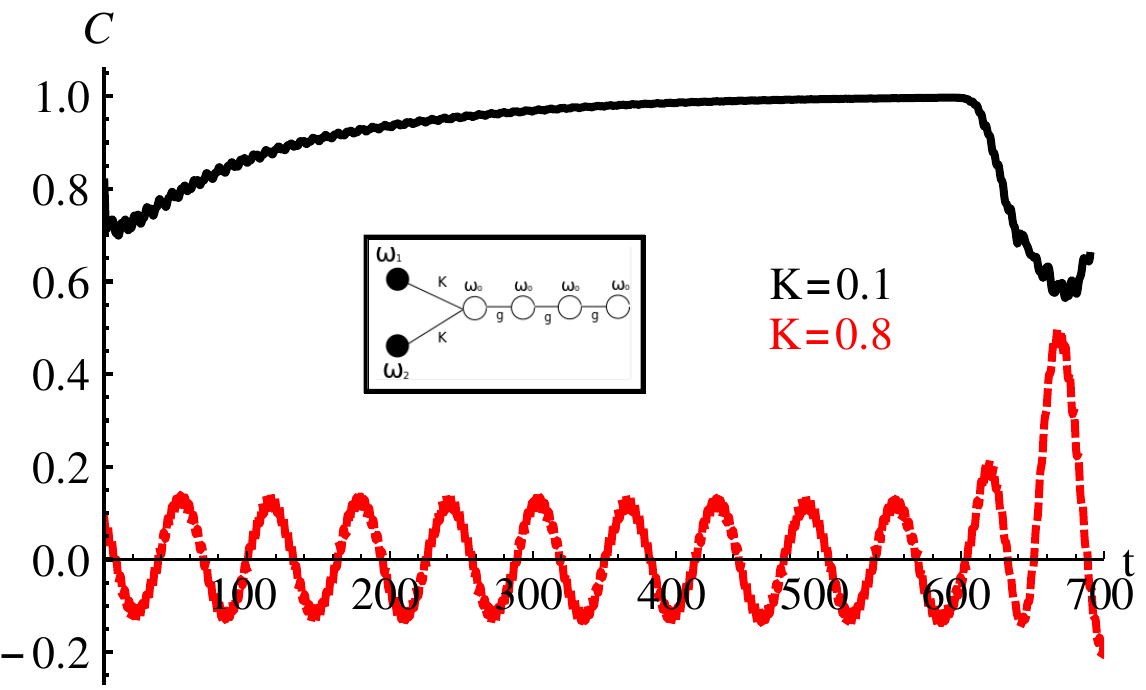} 
\includegraphics[width=0.23\textwidth]{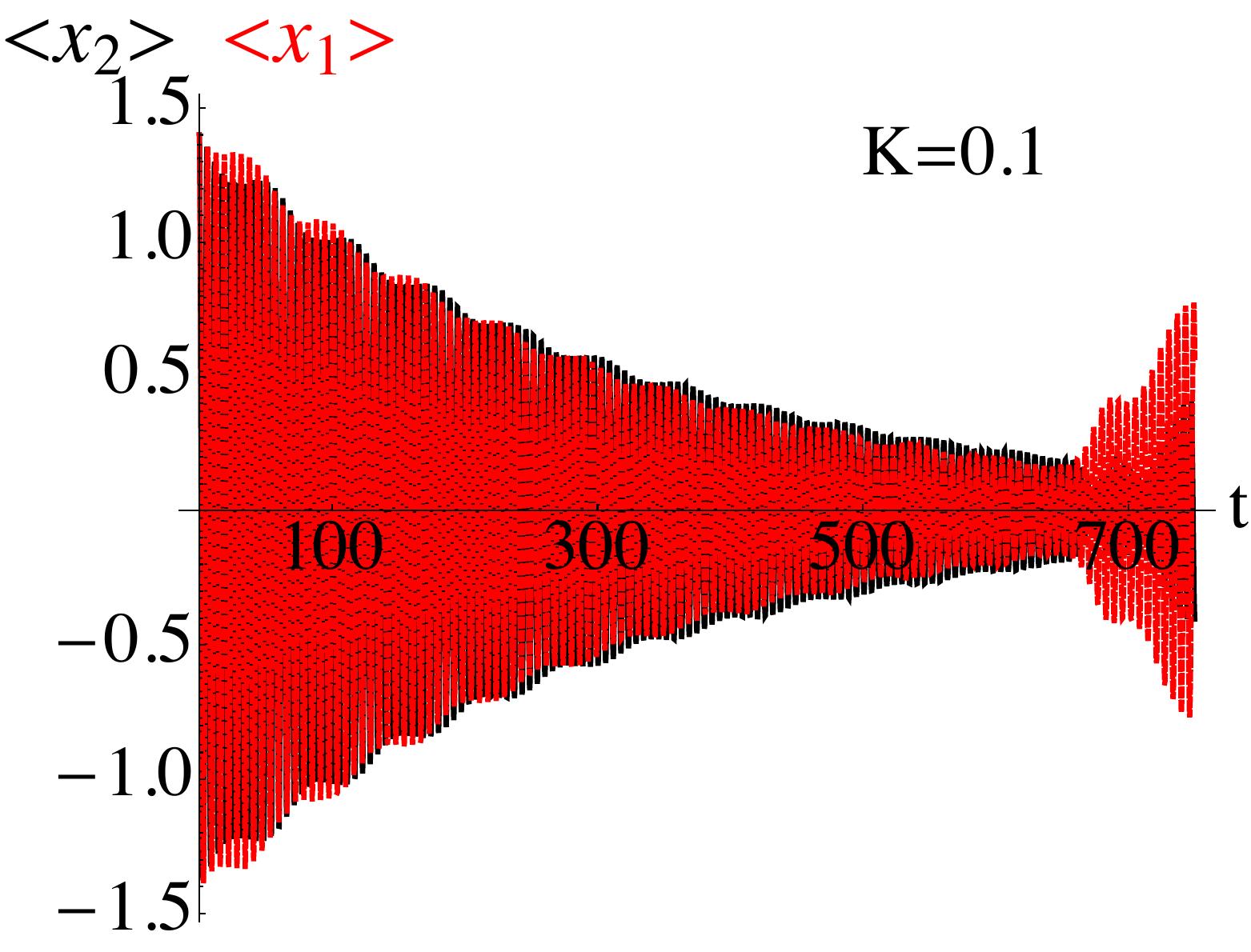} 
\includegraphics[width=0.23\textwidth]{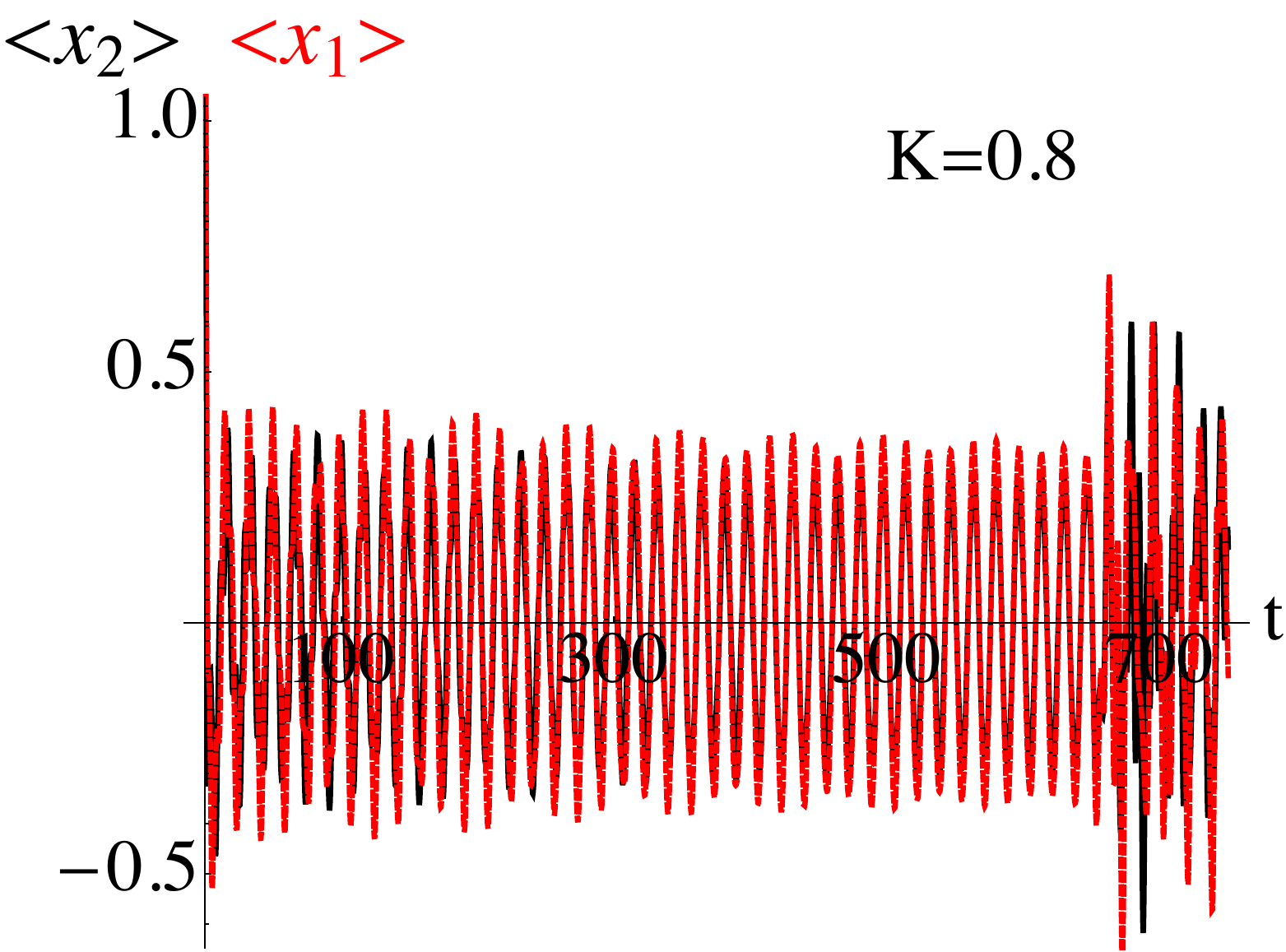} 
\caption{Upper plot: synchronization measure $C(t)$ as a function of
time for two different values of the coupling $K=0.1$ (solid black) and
$0.8$ (dotted red), in unit of $\omega_1^2$.  The other parameters are fixed: 
$\omega_2/\omega_1 = 1.1$, $\lambda = 0$, $M = 300$, $\Omega_0 = 0.4\,\omega_1$, and $g =1.2\,\omega_1^2$.  In the inset, we show a schematic representation of the considered
configuration.  Lower plot: dynamics of the two oscillators plugged
into the first node of the chain, with $K=0.1$ (left panel) and
$K=0.8$, in unit of $\omega_1^2$ (right panel).  $\langle x_{1,2}\rangle$ are in units of $\omega_1^{-1/2}$ and time in units of $\omega_1^{-1}$.}
\label{f3}
\end{figure}

This is not the case when $K\lesssim\omega_i^2$.  We first notice that
this rather large dissipation does not completely deplete the system
energy. Indeed, after a fast transient oscillatory decay, the system
achieves a steady regime of rather large oscillations with constant
amplitude that last up the revival time. $\tau_{R}\simeq 600/\omega_1$ for the
choice of parameters of Fig. \ref{f3}.  In this regime, oscillations are
coherent at frequency smaller than the system frequencies
($0.4<\omega_{1,2}$, Fig. \ref{f3} bottom right) and perfect
synchronization emerges.  After the transmission of the initial pulse,
originated at the edge of the chain, the stiff coupling ($K=0.8\,\omega_1^2$ in
Fig.\ref{f3}) to the environment leads to a steady state in which the
system vibrates at the lowest frequency of the chain, $\Omega_0=0.4\,\omega_1 $.
For decreasing coupling the system depletion of energy continues until
complete damping, whereas for weaker coupling (e.g. $K=0.1\,\omega_1^2$) the system
shows under-damped oscillation at the detuned (Lamb shifted) natural
detuned frequencies, so no synchronization is established.  This
scenario of synchronization occurs for uncoupled probes stiffly attached
to one node of a network until signal reflections (depending on the
network topology) drive the system  away from coherent oscillation, as
shown here at the revival time.
\subsection{Synchronization by cross-talk}
\label{secSRopposite}
\begin{figure}[ht!]
\centering
\includegraphics[width=0.38\textwidth]{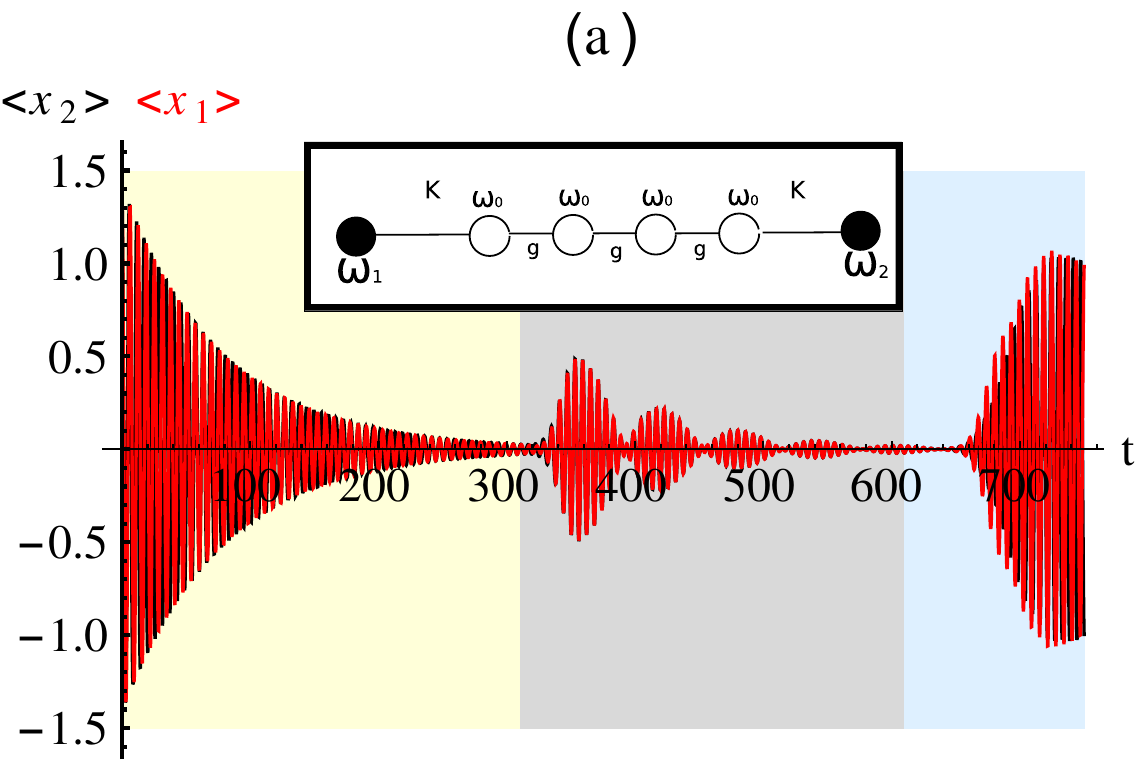}
\includegraphics[width=0.38\textwidth]{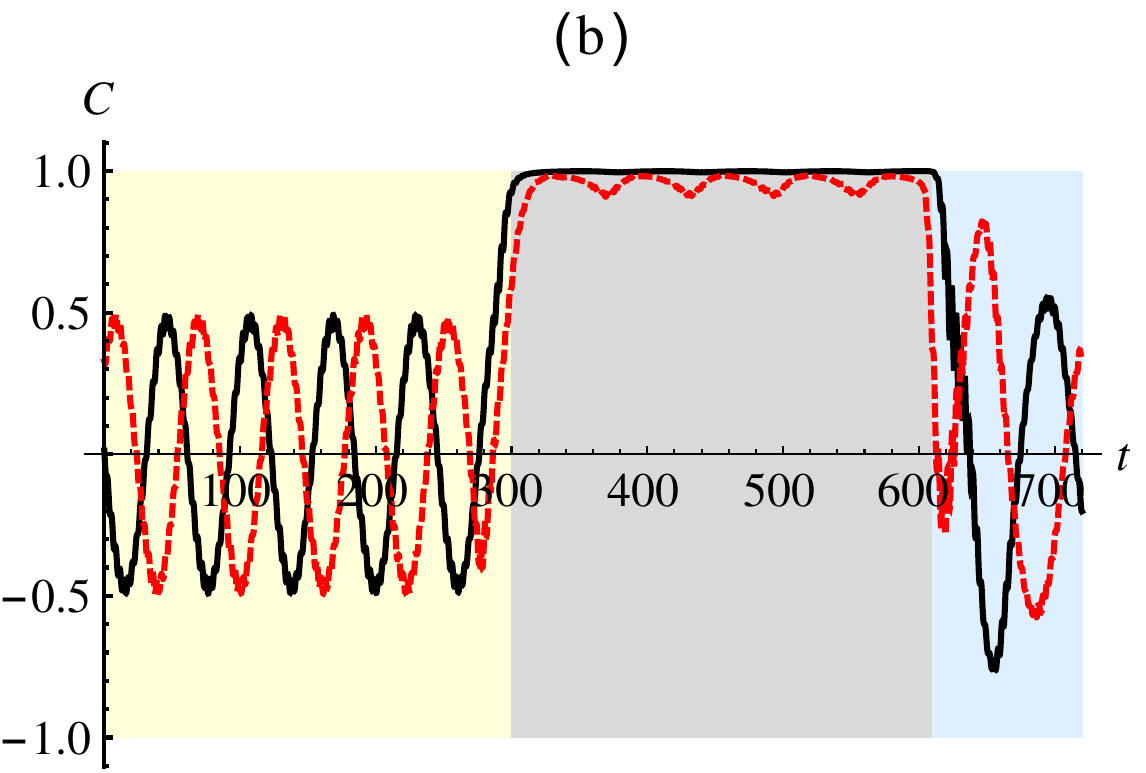} 
\includegraphics[width=0.38\textwidth]{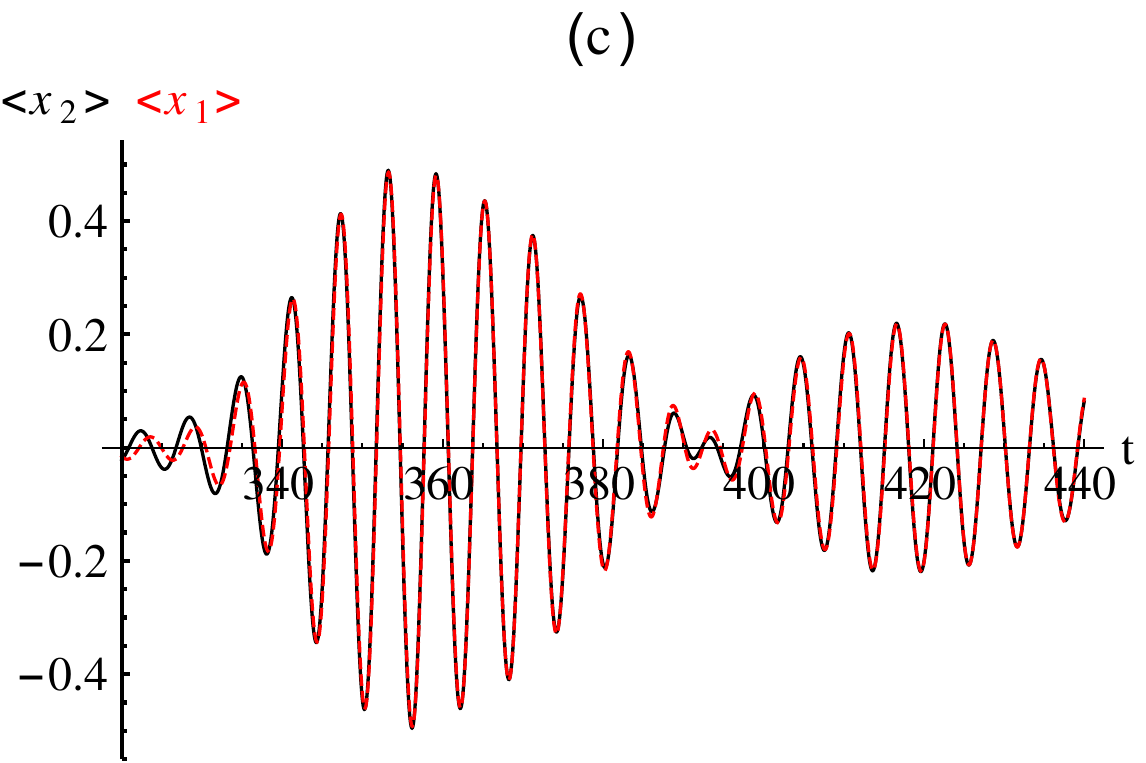} 
\caption{(a) Dynamics, (b) synchronization measure $C(t)$ and (c) zoomed
view of the dynamics for two uncoupled oscillators ($\lambda=0$), with
detuning $\delta\omega/\omega_1=0.1$ coupled to the extremities of the oscillator
chain with coupling constant $K=0.2\,\omega_1^2$ and $M = 300$, $\Omega_0 = 0.4\,\omega_1$ and
$g = 1.2\,\omega_1^2$. The inset in plot (a) shows a schematic representation of the
configuration under analysis.  Plot (b) shows the synchronization
measure for two different initial conditions: symmetric
$x_1(0)=x_2(0)=1.4\,\omega_1^{-1/2}$ and $p_1(0)=p_2(0)=0$ (solid black line) and
asymmetric $x_1(0)=x_2(0)=2\,\omega_1^{-1/2}$ and $p_1(0)=0$, $=p_2(0)=10\,\sqrt{\omega_1}$ (dashed, red
line).  Plot (c) shows the dynamics of the two oscillators $\langle
x_1(t)\rangle$ (solid black line) and $\langle x_2(t) \rangle$ (dashed
red line),  for a time window where synchronization appears. The
background colors in plot (a) and (b) are inserted as a guide for the
eye to mark the three region of independent dissipation (pink),
cross-talk (blue) and revival (light blue).  $\langle x_{1,2}\rangle$ are in units of $\omega_1^{-1/2}$ and time in units of $\omega_1^{-1}$. }
\label{f4}
\end{figure}

The phenomena described in Sect. \ref{secDiss} and \ref{secSRedge} 
show how boundary effects are often detrimental for
synchronization. A different dynamics, however, may take place if 
the two oscillators are allowed to exchange their energy across 
the system. To illustrate this effect we consider a configuration 
where the oscillators do not interact directly, i.e. $\lambda=0$ and 
are plugged at the opposite edges of a 
chain, see Fig.\ref{f4}. 
During an initial transient, even if the probes 
are attached to the same environment (the chain), 
there are not decoherence free sub-spaces \cite{CBSB}: 
the probes actually experience independent dissipation, 
the $\eta$ kernel vanishes, and $x_+$ and $x_-$ are 
coupled to orthogonal modes of the chain.
During this transient the system oscillators lose 
energy and do not synchronize, as shown in  
Fig.\ref{f4}(a) and (b). This is consistent 
with previous studies with infinite and 
separate environments \cite{sync2012}. 
\par
In the weak coupling regime, the undamped 
oscillators start to feel the effect of each other 
after the time interval needed for signal 
propagation through the system, but their 
dynamics still remain incoherent. On the other hand, for 
stronger dissipation $K\lesssim\omega_{1,2}^2$ 
a sudden rise-up of the system oscillations 
appears at the cross-talk  time $\tau_{CT}=\tau_R/2$ 
and perfect synchronization emerges. This behavior is 
illustrated in Fig. \ref{f4}, in the central regions 
of panels (a) and (b).  For $t>\tau_R$ each oscillator 
starts receive its own feedback and 
synchronization is lost again, since it 
is driven back  to its natural frequency.
\par
The mechanism of synchronization found in 
this regime consists of a reciprocal driving 
force of the two system HOs after their local damping: 
at  $\tau_{CT}$ they have lost their initial energy 
due to their dissipation into the cold chain 
($K=0.2\,\omega_1^2$ in Fig.\ref{f4}), and in the cross-talk 
time window $\tau_{CT}< t< \tau_R$ they 
receive a signal from the opposite (detuned) 
system oscillator. A driving at the frequency of 
the opposite oscillator, being detuned, would not 
cause any synchronization, but actually the 
exchanged signals are not at a single frequency,  
having a broad bandwidth due to the transmission 
through the chain. We find that, within the 
cross-talk time window, the oscillator $1$ is 
driven by a signal containing both the main 
frequency component $\omega_2'$ \cite{primes} and 
the resonant one $\omega_1'$ (present in the 
broad signal transmitted through the chain),
leading to the beating signal observed in  
Fig.\ref{f4}(c). A similar scenario occurs 
for the other oscillator $2$, now with the 
strongest and resonant frequency component 
exchanged. Therefore the oscillators placed at the edges, during
cross-talk time, experience a driving force at a signal with the two
detuned frequencies, leading to the characteristic beating signal of
Fig.\ref{f4} and to perfect synchronization. 
\par
This mechanism of synchronization for cross-talk is based on a
reciprocal effect between the system components and is robust when
breaking the symmetry in the initial conditions, even though in this
case $delayed$ synchronization arises. Indeed for non-identical initial
states $\{x_1(0),p_1(0)\}$ and  $\{x_2(0),p_2(0)\}$, the respective
signals  experience a relative phase delay.  Time delay can be taken
into account considering  the delayed signals $\langle x_1(t)\rangle$
and $\langle x_2(t+\Delta t)\rangle$ in $C$. Since this synchronization
scenario is very sensitive to the initial conditions, 
both anti-synchronization and synchronization may arise.
\section {Synchronization and quantum correlations}
\label{secQ}
An interesting question to answer is whether the emergence of
synchronization is accompanied by an increase in the quantum or
classical correlations in the system and if the mechanism of
synchronization by dissipation (Sect. \ref{secDiss}) may be a witness
for the appearance of robust quantum correlations and 
entanglement between the oscillators
pair. As first reported in \cite{sync2012} starting from an entangled
state weakly dissipating into the environment, decoherence and
deterioration of quantum correlations will be reduced in presence of
synchronization.  This mechanism has been analysed also for 3
oscillators \cite{syncLalo} and in networks \cite{syncSR}.
\par
Here we consider instead the possibility to {\em create} correlations and
entanglement starting from product states of the system oscillators and
in relation with quantum synchronization. To this purpose we consider
uncoupled system oscillators starting from an uncorrelated (product)
state with local squeezing.  For identical oscillators
$\omega_1=\omega_2$, entanglement mediated by the reservoir chain and
its dynamical (sudden-death and revival) features have been predicted in
Refs. \cite{morigiY, morigiEPL} in symmetric models.  The possibility to
entangle two oscillators due to strong dissipation in a common bath was
addressed in \cite{correa} while in \cite{syncSR} the case of
dissipative network was treated.  The question we are interested here is
the possibility to create entanglement due to the coherent energy
transmission across the environment between detuned oscillators and in
relation with spontaneous synchronization.  The cases of interest are
for system coupling mediated by the chain ($\lambda=0$)
\begin{figure}
 \centering
 \includegraphics[width=0.45\textwidth]{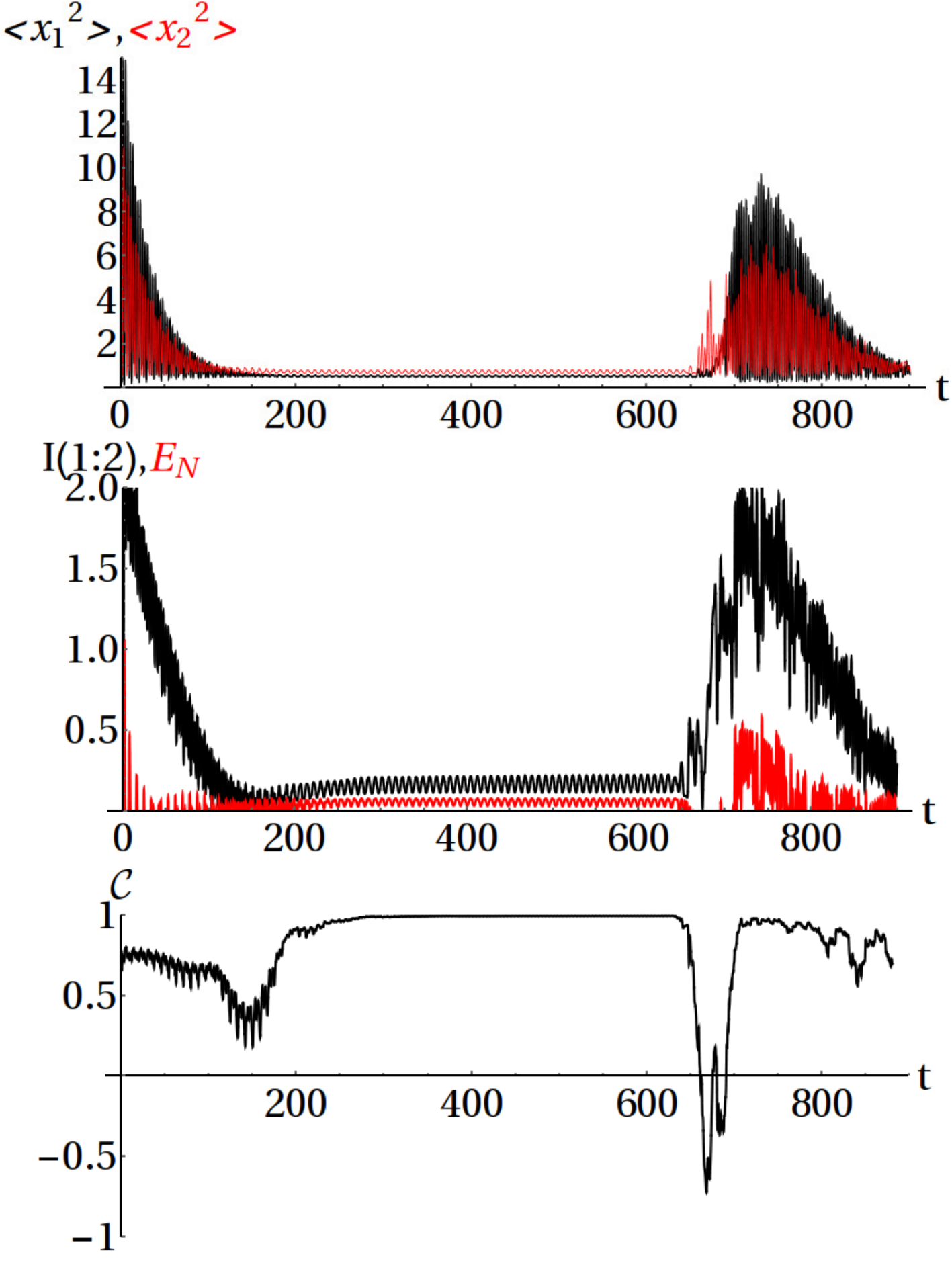} 
  \caption{(From top to bottom) Quadratures, mutual information and
  entanglement, and synchronization factor, starting from a separable,
  squeezed state with squeezing parameters $r_1=r_2=2$, frequencies
  $\omega_2/\omega_1=1.2$ and coupling $\lambda=0$. The two
  oscillators are strongly coupled to the reservoir $K=0.8\,\omega_1^2$, and bath
  parameters are as in previous figures.   $\langle x_{1,2}^2\rangle$ are in units of $\omega_1^{-1}$ and time in units of $\omega_1^{-1}$.}
 \label{fentY}
\end{figure}
We monitor the system entanglement given by the logarithmic negativity
$\mathcal{E}=\max(0,-\ln \nu)$, with $\nu$ the smallest symplectic
eigenvalue of the partially transposed density matrix
\cite{horodecki_review}.  Further we consider the mutual information
$\mathcal{M}= S_A+S_B-S_{AB}$ with $S_i$ Von Neumann entropy of the
reduced system $i=1,2$ and $S_{AB}$ the total entropy.  Actually the
latter has been also suggested to be an order parameter for quantum
synchronization \cite{fazio2}.
\subsection{Coupling to a common chain node}
For a system plugged at the same point of a chain we have seen that two
otherwise uncoupled oscillators ($\lambda=0$) can synchronize in the
strong coupling regime due to the mediating effect of the environment
(see Sect.\ref{secSRedge}).  We find that this synchronization scenario
is also present for system oscillators in vacuum squeezed states.  In
this case synchronization arises between the second order moments, as
shown in Fig. \ref{fentY}.  As for the case of average positions,
fluctuations synchronization is allowed by the strong dissipation, Fig.
\ref{fentY}a, and it later (at $\tau_{R}$) deteriorates due to feedback
effects.  Initially both MI and entanglement are established between the
decoupled and initially uncorrelated system oscillators, as expected,
due to their strong coupling mediated by the chain and the initial local
squeezing, Fig. \ref{fentY}b.  After a transient oscillatory decay
$\mathcal{E}$ and  $\mathcal{M}$ both reach a steady $non-vanishing$
value, consistently with predictions in Ref.\cite{correa}. Here also
synchronization appears, as shown in Fig. \ref{fentY}c, and actually
witnesses entanglement. 
\begin{figure}
 \centering
 \includegraphics[width=0.45\textwidth]{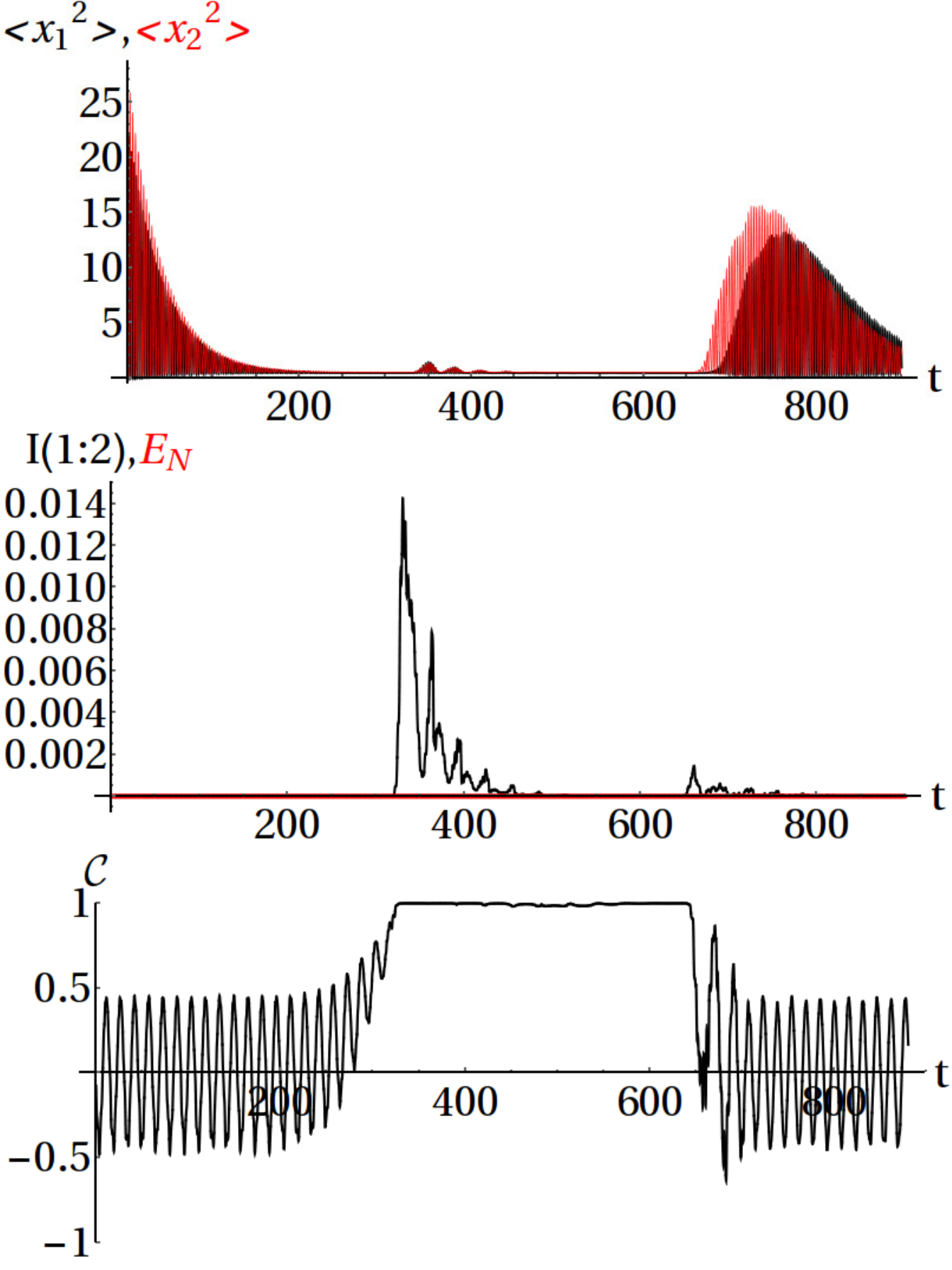} 
  \caption{(From top to bottom) Position variances (top), mutual
  information and entanglement (middle), and synchronization factor
  (bottom), starting from a separable, squeezed state with squeezing
  parameters $r_1=r_2=2$, frequencies $\omega_2/\omega_1=1.2$ and
  coupling $\lambda=0$. The coupling to the reservoir is $K=0.2\,\omega_1^2$, and
  bath parameters are as in previous figures. $\langle x_{1,2}^2\rangle$ are in units of $\omega_1^{-1}$ and time in units of $\omega_1^{-1}$. }\label{fentOr}
\end{figure}
This microscopic model shows that strong coupling to a common
environment allows to  synchronize and entangle uncoupled detuned
oscillators whose interaction is mediated by the environment. This would
persists for an infinite bath while feed-back effects in a finite model
case hinder synchronization, Fig. \ref{fentY}c for $t>\tau_R$.  Further
we notice that the increase of mutual information at $\tau_{R}$ does not
always reflect dynamical synchronization that at the contrary can decay
($\mathcal{C}\simeq 0$ for $650<\omega_1t<700$).
\par
The fact that two uncoupled oscillators, interacting with a chain,
evolve into a synchronized and entangled state is a distinctive effect
of strong coupling, which is not present for weak coupling.
\subsection{Coupling to the chain edges}
We now consider the case in which the oscillators are far-apart, at the
opposite edges of the chain as in Sect. \ref{secSRopposite}.  Is it
possible to synchronize their quantum fluctuations and entangle them due
to the cross-talk?  We consider again  squeezed vacuum states, observing
that the system probes at the opposite edges of the chain evolve toward
a quantum synchronized state in its fluctuations, with a build-up of
correlations during the cross-talk time, as shown by their mutual
information rising from vanishing to finite values (Fig. \ref{fentOr}).
Nevertheless, for reasonable values of the initial squeezing,
entanglement is never created, independently on the initial squeezing
strength. For distant probes therefore synchronization may emerge when
in a cross-talk regime where they exchange energy but this does not lead
to entanglement. 
\section{Conclusions}\label{out}
In conclusion, we have addressed synchronization of two 
quantum oscillators within a finite linear system, and 
analyzed in details the possible
mechanisms leading to a coherent dynamics of the (detuned) system
components. { Besides, we have analyzed in some details 
the connections of  synchronization with the build up of 
entanglement starting
from uncorrelated states and without any direct coupling between the 
two oscillators.}
\par
Our microscopic description has allowed 
us to go beyond the weak dissipation
limit, showing that in the strong coupling regime
new synchronization mechanisms appears among uncoupled 
oscillators, leading to coherent dynamics enabled 
by the environment. Furthermore, cross-talk effects may have a
constructive role, inducing synchronization mediated by signal
transmission. More in general, a condition for spontaneous
synchronization of linear oscillators has been discussed in the context
of the Rayleigh model for vibrations physics.
\par
As a matter of fact, synchronization is important in different contexts but
not always desirable.  For quantum networks
\cite{syncSR}, quantum synchronization witnesses the presence of quantum
correlations which are more robust against dissipation, and even the 
appearance of noiseless sub-systems.  On the other hand, in the context 
of physics of
vibrations, the fact that some vibration modes are damped out very slowly 
may compromise the stability of complex structures \cite{vibrations}.  
\par
Our results pave the way to the analysis  of local synchronization 
mechanisms for small clusters within a larger network, and to
applications of interest for quantum technology and metrology, 
e.g. the use of spontaneous
synchronization to witness quantum correlations or the synchronization of
clocks by coherent coupling.
\begin{acknowledgments}
This work has been supported by EU through the Collaborative 
Project QuProCS (Grant Agreement No. 641277), by MINECO (Grant
No. FIS2014-60343-P), by the "Vicerectorat dâ Investigaci\'o
 i Postgrau"  of the UIB and by the UIB visiting 
 professors program. 
\end{acknowledgments}
\appendix
\section{Hamiltonian normal modes and Langevin equation}\label{app1}
Let's consider a system of two quantum harmonic oscillators,
characterized by frequencies $\omega_1$ and $\omega_2$ and coupling
strength $\lambda$ between them.  The oscillators are plugged with
strength $K$ into an homogeneous chain of quantum HOs at frequency
$\Omega_0$ and chain stiffness $g$.  We now introduce the notation used
to describe the system and environment normal modes (NM):
\begin{align}
H_S&=\frac{\tilde{p_1}^2}{2}+\frac{\tilde{p_2}^2}{2}+\frac{1}{2}\Lambda^2_-q_1^2+\frac{1}{2}\Lambda^2_+q_2^2\\
H_E&=\sum_{j=1}^M\left[\frac{\tilde{P}_j^2}{2}+\frac{1}{2}\Omega^2_jQ_j^2\right]\\
H_{I}&=q_1\sum_{j=1}^M c_{1}(j)Q_j+ q_2 \sum_{j=1}^M c_{2}(j)Q_j.
\end{align}
where the  positions normal modes operators are denoted by $q_{1,2}$ for the system and $Q_j$ for the environment and are computed as:
\begin{align}
 q_1&=\cos\theta\;x_1+\sin\theta\;x_2\\
 q_2&=-\sin\theta\;x_1+\cos\theta\;x_2\\
 X_j&=\sqrt{\frac{2}{M+1}}\sum_{k=0}^M\sin\left(\frac{\pi kj}{M+1}\right) Q_k
 \end{align}
 where the quantity $\theta$ is defined by the relation $$\tan(2\theta)=\frac{2\lambda}{\omega_2^2-\omega_1^2}.$$
The eigenfrequencies and the coupling coefficients are given by:
\begin{align}
 &\Lambda^2_s= \lambda +\frac{\omega_1^2+\omega_2^2}{2}\,+\frac{(-1)^s}{2}\sqrt{4 \lambda ^2+\left(\omega_1^2-\omega_2^2\right)^2}
 \\
 &\Omega^2_j=\Omega_0^2+4g\sin^2\left(\frac{\pi j }{2(M+1)}\right)\\
 & c_{1}(j)=\sqrt{\frac{2K^2}{M+1}}\;\;\times\nonumber\\
 &\hspace{0.7cm} \left[\cos\theta\sin\left(\frac{\pi jm}{M+1}\right)+\sin\theta\sin\left(\frac{\pi jn}{M+1}\right)\right]
 \label{c1} \\
  & c_{2}(j)=\sqrt{\frac{2K^2}{M+1}} \;\;\times\nonumber\\
  &\hspace{0.7cm}\left[\cos\theta\sin\left(\frac{\pi j n}{M+1}\right)-\sin\theta\sin\left(\frac{\pi jm}{M+1}\right)
  \right]\label{c2}.
\end{align}
The dynamics of the system is described by a generalized quantum Langevin equations (GQLE) for operators $q_1(t)$ and $q_2(t)$,
obtained starting from the set of Heisenberg equations for system and environment operators $\{q_{1(2)},\tilde{p}_{1(2)},Q_j,\tilde{P}_j\}$.
The GQLE are integro-differential equations that describe the dynamics of 
the NM operators as a function of the environment parameters and coupling constants:
\begin{align}
&\ddot{q}_1(t)  +[\Lambda^2_{1}-\gamma_{1}(0)] q_{1}(t) \nonumber \\ 
& + \int_0^t dt'[\gamma_{1}(t-t')
\dot{q}_{1}(t')+\eta(t-t')\dot{q}_{2}(t')] \nonumber \\ 
& =-\xi_1(t)
-\gamma_{1}
(t) q_{1}(0)-\eta(t) q_{2}(0) +\eta(0) q_{2}(t)
\label{lang}
\end{align}
and an equivalent expression is found for operator $ q_2$ by replacing
$1\longleftrightarrow 2$.  The kernels $\gamma$ and $\eta$ take the
expressions:
\begin{align}
\gamma_s&(t-t')=\sum_{j=1}^M
\frac{c_s^2(j)}{\Omega_j^2}\cos[\Omega_j(t-t')]\Theta(t-t')\\
\eta&(t-t')=\nonumber\\
&=\sum_{j=1}^M \frac{c_{1}(j)c_{2}(j)}{\Omega_j^2}\cos[\Omega_j(t-t')]\Theta(t-t')
\end{align}
where $s=1,2$ and the external force operator $\xi_{1(2)}(t)$ depends
upon the environment initial conditions \begin{align}
&\xi_s(t)=\nonumber\\
&=\sum_{j}c_s(j)\left(Q_j(0)\cos(\Omega_jt)+\frac{P_j(0)}{\Omega_j}\sin(\Omega_jt)\right)
\end{align}
and gives a zero contribution when averaged over the vacuum state of the
environment.
\begin{figure}
\centering
\includegraphics[width=0.5\textwidth]{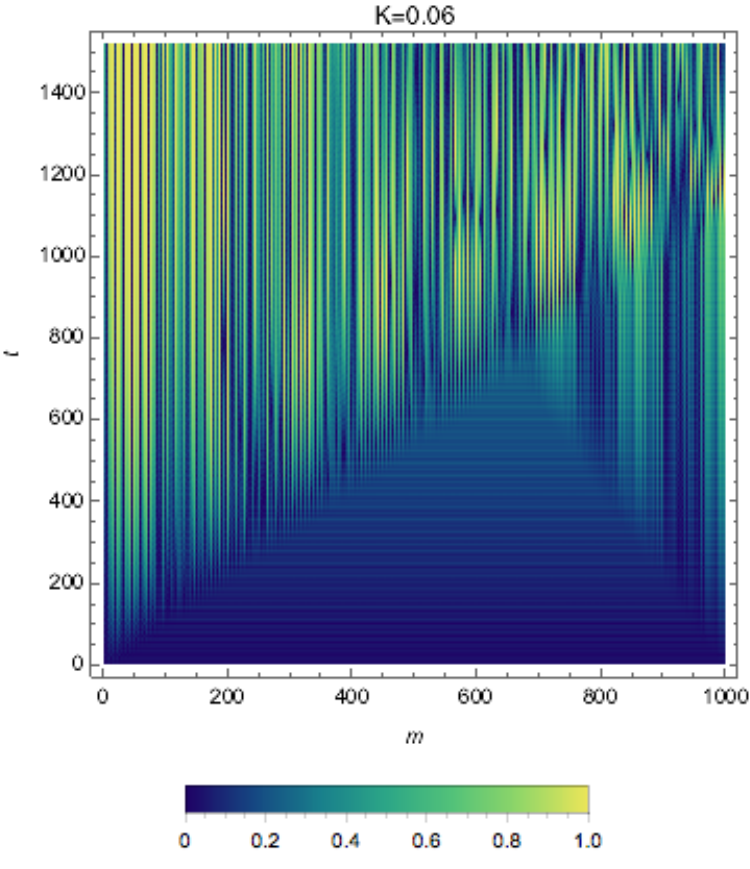} 
\caption{Density plot of the synchronization measure $C(t)$ for a 300-site
chain as a function of the plugging site $m$ and the  time $t$. The
parameters are set as: $\omega_2/\omega_1=1.1$, $\lambda=0.5\,\omega_1^2$,
$K=0.06\,\omega_1^2$, $\Omega_0=0.4\,\omega_1$ and $g=1.2\,\omega_1^2$.} \label{fgeneric}
  \end{figure}
\section{Synchronization across the chain: case $0\ll m\ll N$}\label{app2}
When moving the second oscillator through the chain but far from the
edges, i.e. $0\ll m\ll N$, there is an initial time transient in which
the oscillators do not synchronize and this behavior is independent on
the position  $m$, as shown in Fig.\ref{fgeneric}. This occurs only
before cross-talk and feedback from the boundaries take place and
actually corresponds to a good approximation of independent dissipation
of the two detuned systems.  In particular, if the coupling is weak and
time is long enough to have only the resonant system-bath interaction
surviving, then one expects that the chain normal modes $Q_k\pm$ that
are resonant with the system eigenfrequencies dominate the dynamics
($k_-$ resonates with the system normal mode $q_1$ and $k_+$ with $q_2$)
 leading to an effective interaction
\begin{align}
 H_{I}=K\sqrt{\frac{2}{M+1}}\left[c_1 q_1Q_{k_-}+c_2q_2Q_{k_+}\right].
\end{align}
with
\begin{align}
 c_1=\left[\cos\theta\;\sin\left(\frac{\pi
 k_-}{M+1}\right)+\sin\theta\;\sin\left(\frac{\pi k_-
 m}{M+1}\right)\right]\\ c_2=\left[\cos\theta\;\sin\left(\frac{\pi k_+
 m}{M+1}\right)-\sin\theta\;\sin\left(\frac{\pi k_+}{M+1}\right)\right]
\end{align}
Nevertheless, such effective resonant interaction is established after
long times, while during the transient analyzed here there are bands of
normal modes that are exchanging energy with the system.  The average
effect of several of such modes of the chain leads to 
$c_1\sim \cos\theta\;\sin\left(\frac{\pi k_-}{M+1}\right)$ and
$c_2\sim\sin\theta\;\sin\left(\frac{\pi k_+}{M+1}\right)$, so that the
system normal modes decay at some rate independently on their distance
$m$, as shown in the triangular region in Fig. \ref{fgeneric}.
The spatio-temporal synchronization diagram  shown in Fig.
\ref{fgeneric} clearly displays the effects of cross-talk and
reflections from the boundaries, leading to a strong and non-monotonic
dependence on $m$ and, often, to synchronization for larger times.


\begin{thebibliography}{0}
\bibitem{clmode} Here `classical' is the adopted terminology used in vibration physics, not to be confused with the 
absence of quantumness.
\end{thebibliography}
\end{document}